\documentclass[9pt, conference]{IEEEtran}
\IEEEoverridecommandlockouts
\usepackage{cite}
\usepackage{amsmath,amssymb,amsfonts}
\usepackage{graphicx}
\usepackage{textcomp}
\usepackage{xcolor}
\usepackage{multirow}
\usepackage{hyperref}
\usepackage{algorithm,algpseudocode}
\usepackage{amssymb}

\def\BibTeX{{\rm B\kern-.05em{\sc i\kern-.025em b}\kern-.08em
    T\kern-.1667em\lower.7ex\hbox{E}\kern-.125emX}}
\begin{document}

\title{SegAug: CTC-Aligned Segmented Augmentation For Robust RNN-Transducer Based Speech Recognition \\
% {\footnotesize \textsuperscript{*}Note: Sub-titles are not captured for https://ieeexplore.ieee.org  and
% should not be used}
% \thanks{Identify applicable funding agency here. If none, delete this.}
}
\author{\IEEEauthorblockN{Khanh Le\IEEEauthorrefmark{1}, Tuan Vu Ho, Dung Tran\IEEEauthorrefmark{2} and Duc Thanh Chau\IEEEauthorrefmark{3}\IEEEauthorrefmark{4}}
\IEEEauthorblockA{\IEEEauthorrefmark{1}ZaloAI, Vietnam}
\IEEEauthorblockA{\IEEEauthorrefmark{2}Independent Researcher}
\IEEEauthorblockA{\IEEEauthorrefmark{3}Ho Chi Minh City University of Science \\}
\IEEEauthorblockA{\IEEEauthorrefmark{4}Vietnam National University, Ho Chi Minh City
}
\texttt{\{khanhld218, tuanvu.ksvt92, dung.n.tran.xyz\}@gmail.com, ctduc@fit.hcmus.edu.vn} \\

\textit{Corresponding author: Duc Thanh Chau}

}

\maketitle

\begin{abstract}
RNN-Transducer (RNN-T) is a widely adopted architecture in speech recognition, integrating acoustic and language modeling in an end-to-end framework. However, the RNN-T predictor tends to over-rely on consecutive word dependencies in training data, leading to high deletion error rates, particularly with less common or out-of-domain phrases. Existing solutions, such as regularization and data augmentation, often compromise other aspects of performance. We propose SegAug, an alignment-based augmentation technique that generates contextually varied audio-text pairs with low sentence-level semantics. This method encourages the model to focus more on acoustic features while diversifying the learned textual patterns of its internal language model, thereby reducing deletion errors and enhancing overall performance. Evaluations on the LibriSpeech and Tedlium-v3 datasets demonstrate a relative WER reduction of up to 12.5\% on small-scale and 6.9\% on large-scale settings. Notably, most of the improvement stems from reduced deletion errors, with relative reductions of 45.4\% and 18.5\%, respectively. These results highlight SegAug's effectiveness in improving RNN-T's robustness, offering a promising solution for enhancing speech recognition performance across diverse and challenging scenarios.
\end{abstract}

\begin{IEEEkeywords}
segaug, speech recognition, rnn-transducer
\end{IEEEkeywords}

\section{Introduction}
\label{sec:intro}
% - \textbf{intro to the problem}\\
Automatic Speech Recognition (ASR) systems have seen remarkable advancements with the advent of Recurrent Neural Network Transducers (RNN-T). These models have become a cornerstone in various applications, from virtual assistants to transcription services. However, despite their success, RNN-T models commonly suffer from high deletion error rates, especially when processing low-semantic-content phrases or out-of-domain inputs \cite{MWER, kim22f_interspeech, 9746948}. This is due to the context dependency directly embedded within the posterior distribution, as RNN-T models implicitly learn an internal language model (ILM) as a sequence prior restricted to the audio transcription only. These deletions can significantly degrade the overall accuracy of ASR systems, often resulting in missing words that are critical to the intended meaning. In dictation applications, deletion errors are especially problematic, as users speak without real-time text response, and no matter how many times they try to spell the word, it fails to appear, leading to a frustrating user experience.
% - \textbf{review the pertinent literature: related works - the gaps, good and bad}\\

Existing approaches to mitigate deletion errors, such as the integration of external language models \cite{MWER} or the use of sparse attention mechanisms \cite{kim22f_interspeech}, have shown some success but often come with increased computational complexity and limited scalability. Some efforts, such as \cite{9746948, 9053600}, attempt to estimate the impact of the ILM learned by RNN-T's predictor and then correct it in the posterior during decoding to integrate an external language model, aiding in generalization to out-of-domain data. While these methods are effective, they primarily focus on suppressing the ILM’s influence during the decoding phase. Efforts to diversify or generalize the ILM during training remain limited, leaving room for further exploration in this area.
% - \textbf{identify the gap in the literature that the current research was intended to address.} \\
% - \textbf{method of the investigation.} \\

In this paper, we propose SegAug, a novel augmentation technique specifically designed to reduce deletion errors in RNN-T models, making it a promising augmentation-based approach for addressing this issue.
% - \textbf{principal results} \\
Unlike traditional augmentation methods such as Speed Perturb \cite{ko15_interspeech} or SpecAug \cite{Park_2019}, which primarily focus on improving general robustness, SegAug is specifically designed to address the critical issue of deletion errors in RNN-T models. It starts with a CTC-aligned word segmentation process, followed by the random application of four sub-augmentation techniques—SegDrop, SegPerm, SegCrop, and SegMix. These techniques work together to create diverse, low-semantic-content audio-text pairs, targeting the reduction of deletion errors, a major contributor to the overall Word Error Rate (WER). By recalibrating the model’s acoustic sensitivity and strengthening the textual dependencies within the ILM, SegAug enhances RNN-T performance, making its predictor component more robust, comprehensive, and less prone to bias from training data.

We rigorously evaluate SegAug’s performance across a variety of settings, including small and large-scale datasets, as well as in-domain and out-of-domain scenarios, demonstrating its significant impact on reducing deletion errors and overall WER. To reinforce our findings, we compute confidence intervals using the bootstrapping approach, providing a measure of the statistical reliability of our results. Furthermore, we extend our analysis to other end-to-end ASR models, such as Attention-based Encoder-Decoder (AED) systems, where SegAug continues to demonstrate remarkable effectiveness, highlighting its robustness and versatility across diverse applications.
\section{Internal Language Model In RNN-T}
\label{sec:format}
RNN-T is a neural network architecture designed to handle the variable lengths of input and output sequences in ASR tasks. Proposed by \cite{STWRNN}, RNN-T consists of three main components: an encoder, a prediction network, and a joint network. 
% The encoder processes the input sequence \( \mathbf{x} = (x_1, x_2, \ldots, x_T) \), the prediction network generates predictions based on previous outputs, and the joint network combines the outputs of both to produce the final prediction. 
% \begin{equation}
%     % \mathbf{h}^{\text{joint}}_{t,u} = \text{Joint}(\mathbf{h}^{\text{enc}}_t, \mathbf{h}^{\text{pred}}_u)
% \mathbf{h}^{\text{joint}}_{t,u} = \phi(W_{\text{enc}} \mathbf{h}^{\text{enc}}_t + W_{\text{pred}} \mathbf{h}^{\text{pred}}_u + \mathbf{b})
% \end{equation}
% where \( \mathbf{h}^{\text{enc}}_t \) is the hidden state of the encoder at time step \( t \) and \( \mathbf{h}^{\text{pred}}_u \) is the hidden state of the prediction network at output step \( u \). \( W_{\text{enc}} \) and \( W_{\text{pred}} \) are learned weight matrices applied to the encoder and prediction network outputs, respectively, \( \mathbf{b} \) is a bias term, and \( \phi \) is a non-linear activation function, which is tanh in our case.
% The probability of an output sequence \( \mathbf{y} = (y_1, y_2, \ldots, y_U) \) given an input sequence \( \mathbf{x} \) is defined as:
% \begin{equation}
%     P(\mathbf{y}|\mathbf{x}) = \prod_{u=1}^{U} P(y_u | y_{1:u-1}, \mathbf{x}) =  \prod_{u=1}^{U} \text{Softmax}(W \mathbf{h}^{\text{joint}}_{t,u} + b)
% \end{equation}
% where $W$ and $b$ are learnable parameters.
The encoder processes the input sequence \( \mathbf{x} = (x_1, x_2, \ldots, x_T) \), the prediction network generates predictions based on previous outputs, and the joint network combines the outputs of both to produce the final prediction.
\begin{equation}
\mathbf{h}^{\text{joint}}_{t,u} = \phi(W_{\text{enc}} \mathbf{h}^{\text{enc}}_t + W_{\text{pred}} \mathbf{h}^{\text{pred}}_u + \mathbf{b}_1),
\end{equation}
where \( \mathbf{h}^{\text{enc}}_t \) is the hidden state of the encoder at time step \( t \), and \( \mathbf{h}^{\text{pred}}_u \) is the hidden state of the prediction network at output step \( u \). \( W_{\text{enc}} \) and \( W_{\text{pred}} \) are learned weight matrices applied to the encoder and prediction network outputs, respectively, \( \mathbf{b}_1 \) is a bias term, and \( \phi \) is a non-linear activation function, which is \( \tanh \) in this case.

The probability of an output sequence \( \mathbf{y} = (y_1, y_2, \ldots, y_U) \) given an input sequence \( \mathbf{x} \) is defined as:
\begin{equation}
    P(\mathbf{y}|\mathbf{x}) = \prod_{u=1}^{U} P(y_u | y_{1:u-1}, \mathbf{x}) =  \prod_{u=1}^{U} \text{Softmax}(W \mathbf{h}^{\text{joint}}_{t,u} + \mathbf{b}_2),
\end{equation}
where \( W \) and \( \mathbf{b}_2 \) are learnable parameters.
The model is trained to maximize the log-probability of the correct output sequence:
\begin{equation}
    \mathcal{L}_{\text{RNN-T}} = -\log P(\mathbf{y}|\mathbf{x})
\end{equation}
The RNN-T model employs the predictor, which functions as a language model, incorporating linguistic context into the decoding process and producing more accurate and coherent outputs. However, the integrated language modeling aspect of RNN-T can sometimes lead to over-reliance on the linguistic context of the audio transcript, causing the model to drop words that rarely appear in the surrounding context of the training data, even if they are correct. This is even worse in low-resource training data, which does not sufficiently cover uncommon phrases or rare words.
\section{Method}
\label{sec:pagestyle}
\subsection{SegAug Policies}
\label{subsec:segaug}
 \begin{figure}[ht]
    \vspace{-11pt}
  \centering
  \includegraphics[width=\linewidth]{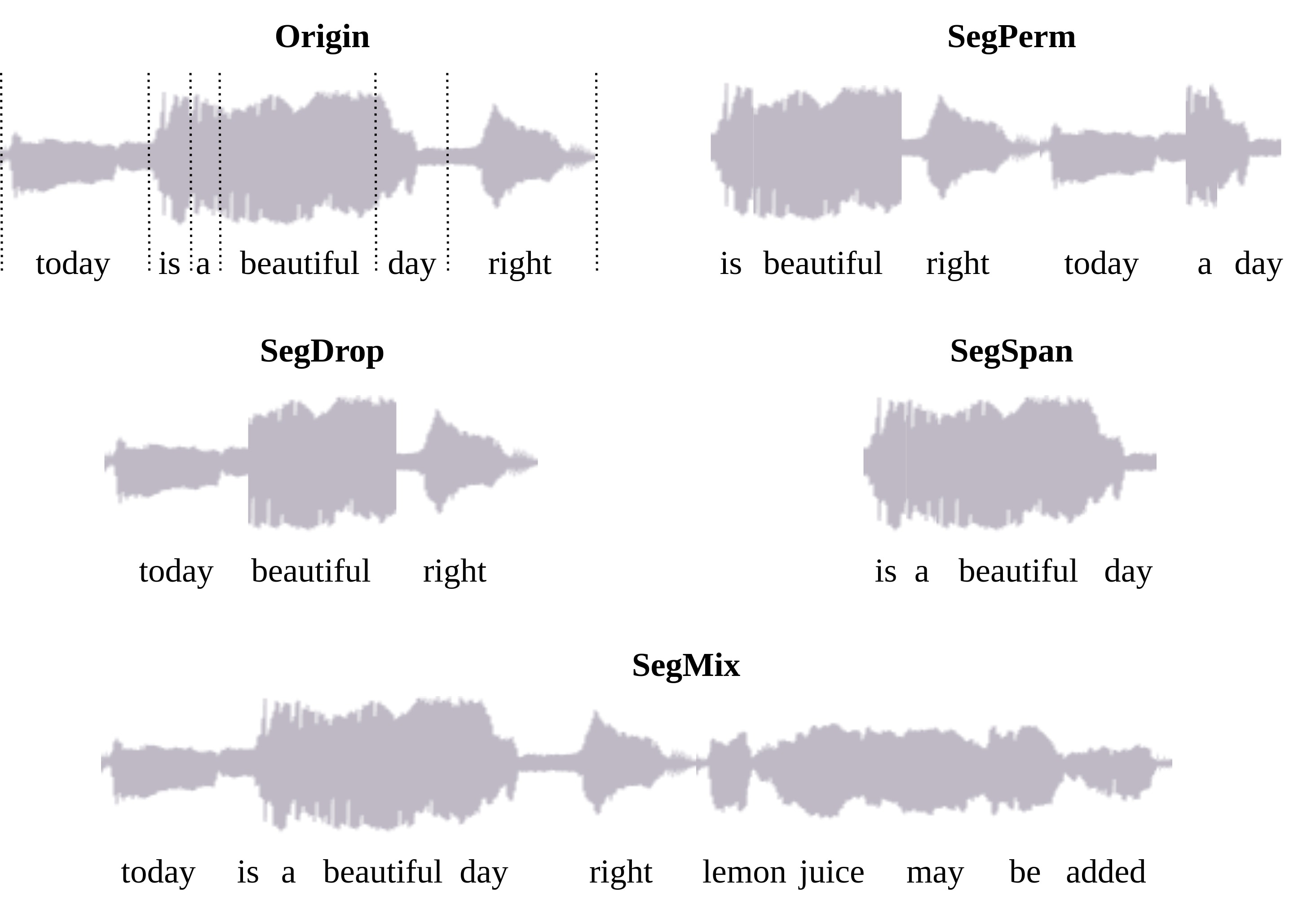}
\vspace{-17pt}

  \caption{Illustration of SegAug transformation methods.}
  \label{fig:figure2}
\vspace{-2pt}
\end{figure}
SegAug includes four sub-augmentation techniques, each designed to tackle different aspects of the problem by generating diverse and complex audio-text pairs. The four sub-augmentation methods, including SegDrop, SegPerm, SegCrop, and SegMix, are illustrated in Figure \ref{fig:figure2}. The SegAug policies are outlined in Algorithm \ref{alg:alg1}, with probability settings of 50\% and 75\% selected through hyperparameter search based on the validation set. Detailed explanations of these settings are provided in the experimental section. \\
\textbf{SegDrop (SD)}: This technique involves randomly removing up to 50\% of the words and their corresponding audio segments from an utterance. By doing so, we simulate scenarios where parts of the speech are missing or obscured, which encourages the model to become more robust to incomplete data and focus on the available acoustic cues. \\
\textbf{SegPerm (SP)}: The process of SegPerm is a combination of \cite{SDAFEMSR} and \cite{specswap}. Specifically, we preserve word-level semantic swapping, unlike the random swapping used in \cite{specswap}, ensuring that the swapped audio and text pairs remain aligned. Furthermore, we overcome the language-specific barrier by removing the swapping rules of \cite{SDAFEMSR} to generate more complex and diverse sequences, pushing the model to focus on the acoustic part while enhancing its robustness to textual variations.\\
\textbf{SegCrop (SC)}: In our production, the RNN-T model often misses words at the beginning of speech. We found that these words are less likely to appear at the beginning of sentences in the training dataset. Therefore, SegCrop randomly chooses one part of the speech and drops the others. Unlike SegDrop, SegCrop mostly preserves the sentence-level meaning while dropping segments.
% \vspace{-2pt}
\begin{algorithm}
\caption{SegAug Policies} \label{alg:alg1}
\textbf{Input}: audio text pair $x$, audio text pair $y$ \\
\textbf{Output}: list of augmented audio text pair $o$
\begin{algorithmic}[1]
\Function{augmenter}{}
    \State \Return $\text{choice}(\text{SC}, \text{SP}, \text{SD}, p = [0.1, 0.6, 0.3])$
\EndFunction
\State $o \gets \text{[]}$
\If{$\text{random()} \leq 0.5$}
    \If{$\text{random()} \leq 0.75$}
        \For{$i \in [x, y]$}
            \State $\text{aug\_i} \gets \text{augmenter}(i)$
            \State $o.\text{append}(\text{aug\_i})$
        \EndFor
    \Else
        \State $\text{xy} \gets \text{SM}(x, y)$
        \State $\text{aug\_xy} \gets \text{augmenter}(\text{xy})$
        \State $o.\text{append}(\text{aug\_xy})$
    \EndIf
\EndIf
\State \textbf{return} $o$
\end{algorithmic}
\end{algorithm} \\
\textbf{SegMix (SM)}: SegMix concatenates two audios and their corresponding transcriptions to form a new audio sequence. However, simple concatenation only affects the context near the join. Therefore, after concatenation, we further randomly apply the three augmentations above to generate more complex audio and text patterns.

SegAug is directly applied to waveform data and executed on the fly within the data pipeline. Word alignments are precomputed using the CTC algorithm, as detailed in Section \ref{subsec:ctc}. Notably, we observed no additional computational overhead from SegAug during the training phase. Existing methods, such as those described in \cite{SDAFEMSR, 10097196}, also build on similar principles but rely heavily on part-of-speech tagging and rules for segment swapping or mixing. While these methods preserve sentence semantics, they face implementation challenges due to multi-step processing and lack of support for on-the-fly training, limiting the diversity of generated sequences.
\subsection{CTC-Aligned Segmentation}
\label{subsec:ctc}
CTC, introduced by \cite{CTCLUSDWRNN}, is an algorithm specifically designed for sequence-to-sequence problems where the alignment between input and output sequences is unknown. CTC introduces a special blank token \( \varnothing \) to handle the alignment problem and the CTC loss function sums over all possible alignments \( \pi \) of the input sequence \( \mathbf{x} \) to the output sequence \( \mathbf{y} \):
\begin{equation}
    P(\mathbf{y}|\mathbf{x}) = \sum_{\pi \in \mathcal{B}^{-1}(\mathbf{y})} P(\pi|\mathbf{x})
\end{equation}
where \( \mathcal{B} \) is a function that removes repeated characters and blank tokens from \( \pi \), and \( \mathcal{B}^{-1}(\mathbf{y}) \) refers to this inverse mapping, representing the set of all alignments that correspond to \( \mathbf{y} \). The CTC loss is then given by:
\begin{equation}
    \mathcal{L}_{\text{CTC}} = -\log P(\mathbf{y}|\mathbf{x})
\end{equation}
We employ the Viterbi algorithm to find the most likely alignment of the transcription sequence given the model's prediction. Let \( \alpha_t^i \) represent the probability of being in state \( i \) at time \( t \). The Viterbi algorithm updates these probabilities as follows:
\begin{equation}
\resizebox{0.9\columnwidth}{!}{$
\alpha_t^i = 
\begin{cases}
\alpha_{t-1}^i \cdot P(x_t = \varnothing) & \text{if } \mathbf{Y}'_i = \varnothing \\
\max (\alpha_{t-1}^i, \alpha_{t-1}^{i-1}) \cdot P(x_t = \mathbf{Y}'_i) & \text{if } \mathbf{Y}'_i = \mathbf{Y}_{i-1} \\
\max (\alpha_{t-1}^i, \alpha_{t-1}^{i-1}, \alpha_{t-1}^{i-2}) \cdot P(x_t = \mathbf{Y}'_i) & \text{otherwise}
\end{cases}
$}
\end{equation}
Here, \( \alpha_t^i \) is the probability of the most likely path ending in state \( i \) at time \( t \), and \( P(x_t = \mathbf{Y}'_i) \) is the probability of observing the output \( \mathbf{Y}'_i \) at time \( t \). We use \( \mathbf{Y}' \) instead of \( \mathbf{Y} \) to refer to the target sequence augmented with blank symbols.
To find the best path, we need to backtrack from the state with the highest probability at the final time step. Let \( \delta_t \) denote the state at time \( t \) in the best path.

1. Initialization:
\begin{equation}
   \delta_T = \arg\max_{i} \alpha_T^i
\end{equation}

2. Recursion:
   For \( t = T-1, T-2, \ldots, 1 \):
\begin{equation}
   \delta_t = 
   \begin{cases}
   \delta_{t+1} & \text{if } \alpha_t^{\delta_{t+1}} \cdot P(x_{t+1} = \mathbf{Y}'_{\delta_{t+1}}) = \alpha_{t+1}^{\delta_{t+1}} \\
   \delta_{t+1} - 1 & \text{if } \alpha_t^{\delta_{t+1}-1} \cdot P(x_{t+1} = \mathbf{Y}'_{\delta_{t+1}}) = \alpha_{t+1}^{\delta_{t+1}} \\
   \delta_{t+1} - 2 & \text{if } \mathbf{Y}'_{\delta_{t+1}} \neq \varnothing \text{ and } \mathbf{Y}'_{\delta_{t+1}} \neq \mathbf{Y}'_{\delta_{t+1}-2}
   \end{cases}
\end{equation}

3. Final Path:
   The best path \( \mathbf{\delta} \) is given by the sequence of states \( \delta_t \) for \( t = 1, 2, \ldots, T \).
   
This process reconstructs the most likely sequence of states that generated the observed sequence, allowing us to determine the best alignment of the input and output sequences. 
In experiments, we find that using word or subword-level alignment models can be problematic due to their larger unit sizes, which can obscure fine-grained temporal details and make it harder for the model to learn precise alignments. Therefore, we use character-level training for the CTC model to align segments of words in the audio. Character-level models can more effectively capture nuances in pronunciation and speech variability, leading to more accurate segmentation. Additionally, because blank frames commonly appear around word boundaries, we refine the segmentation by repositioning these boundaries to the centers between words.
\section{Experiments}
\label{sec:experiments}
\subsection{Setups} \label{sec:3.1}
\noindent\textbf{CTC Model:} The model was implemented using the Wenet toolkit \cite{zhang2022wenet}. The input features were initially processed by a convolutional subsampling layer, consisting of a 2-layer convolutional neural network with 256 channels, a stride of 2, and a kernel size of 3, before being forwarded to the encoder network. The Conformer-based encoder network comprises 12 encoder layers, each with 256 hidden units and 4 self-attention heads, and the feed-forward layer has outputs of 2048 dimensions. \\
% All the CTC models are trained using Speed Perturb \cite{ko15_interspeech} and SpecAug augmentations \cite{Park_2019}. \\
\textbf{RNN-T Model:} The RNN-T encoder is configured identically to the CTC model. The prediction network uses  2-stacked LSTM layers with 256 cells each, and the joint network includes 512 hidden units. \\ 
\textbf{AED Model: } Similarly, the AED model uses the same encoder as the CTC and RNN-T models. The decoder is a bi-transformer with 3 forward and 3 reverse blocks, each consisting of 4 attention heads and a feed-forward layer with 2048 units. \\
\textbf{Dataset:} We conducted experiments on the Librispeech dataset, comprising 960 hours of training audio. The training data was divided into two settings: a small scale with the subset train-clean-100 hours of audio and a large scale with the full training data. Out-of-domain evaluation was carried out using the Tedlium-v3 dataset. \\
\textbf{Training:} Our audio front end utilizes 80-dimensional filter-bank features with 25ms FFT windows and a 10ms frameshift. The Byte-Pair Encoding vocabulary consists of 40 and 5000 tokens for character-level and subword-level, respectively, each associated with a 256-dimensional embedding. All models, except for the alignment CTC model trained at the character level, are trained at the subword level. All CTC and AED models are augmented with Speed Perturb and SpecAug. The training was conducted from scratch on 4 NVIDIA 80GB H100 GPUs with mixed precision, lasting 150 epochs with the Adam optimizer and a Noam warm-up learning rate scheduler, spanning 5000 steps with a peak learning rate of 0.001. The final model averages parameters from the last 75 checkpoints and greedy search is applied for inference. \\
\textbf{Statistical evaluation:} We also compute the confidence interval at a significance level of $\alpha = 5\%$ using the bootstrapping method with $B = 5000$ iterations for the metric of interest, utilizing the toolkit from \cite{Confidence_Intervals}.
The result is represented as $\textit{mean}_{[\textit{min}, \textit{max}]}$, where \textit{mean} refers to the metric value on the original test set and the range denotes the lower and upper bounds of the interval.
\subsection{Results on small scale dataset}
\label{sec:results}
\begin{table*}[ht]
\vspace{-20pt}
\centering
\caption{\label{tab:table1} Experimental results for various models on small and large scale datasets across the LibriSpeech and Tedlium test sets, comparing the effects of Speed Perturb, SpecAug, and SegAug on RNN-T models.}
\vspace{-4pt}
\resizebox{\textwidth}{!}{

    \begin{tabular}{ | c | l | c  c  c | c  c  c | c  c  c | c  c  c |}
        \hline
     \multirow{3}{*}{\textbf{\#}} & \multirow{3}{*}{\textbf{Model}} & \multicolumn{6}{c|}{\textbf{Small scale}}  & \multicolumn{6}{c |}{\textbf{Large scale}} \\
     \cline{3-14}
     & & \multicolumn{3}{c |}{\textbf{WER} (\%)} & \multicolumn{3}{c |}{\textbf{DEL} (\%)} & \multicolumn{3}{c |}{\textbf{WER} (\%)} & \multicolumn{3}{c |}{\textbf{DEL} (\%)} \\
     \cline{3-14}
     & & \textbf{clean}  & \textbf{other} & \textbf{Ted-v3} & \textbf{clean} & \textbf{other} & \textbf{Ted-v3} & \textbf{clean}  & \textbf{other} & \textbf{Ted-v3} & \textbf{clean} & \textbf{other} & \textbf{Ted-v3} \\
     \hline
     % \cline{2-14}
     $C_0$ & \text{Character CTC} & 8.21  & 24.65 & 30.85 & 0.65 & 2.35 & 6.58 & 3.36 & 9.14 & 20.70 & 0.25 & 0.73 & 5.75 \\
     % $C_1$ & \text{Subword CTC} & 8.22 & 22.22 & 29.06 & 0.81 & 2.84 & 7.26 & 3.56 & 8.93 & 20.43 & 0.28 & 0.80 & 5.92 \\
    \hline
     $R_0$ & \text{RNN-T} & 9.04 & 26.83 &	32.64 & 1.46 & 5.27 & 9.98 & 3.77 &	10.43 & 21.94 & 0.38 & 1.22 & 6.49 \\
     $R_1$ & \text{$R_0$ + Speed Perturb} & 8.67 & 24.47 & 31.14 & 1.42 & 4.55 &	9.78 & 3.67 & 10.16 & 21.43 &	0.39 & 1.24 & 6.30 \\
     $R_2$ & \text{$R_0$ + SpecAug} & 8.26 & 22.21 & 29.86 & 1.68 & 5.25 & 10.19 & 3.20 & 8.19 & 20.01 & 0.31 & 1.03 & 6.49 \\
     $R_3$ & \text{$R_2$ + Speed Perturb} & 7.99 & 21.34 & 28.99 & 1.63 & 4.94 &	10.32 & 3.16 & 8.28 & 19.99 & 0.35 & 1.08 & 6.28 \\

     $R_4$ & \text{$R_0$ + SegAug} & 8.39 & 26.10 & 30.30 & 0.99 & 4.08 & 8.59 & 3.54 & 9.46 & 20.52 & 0.35 & 1.06 & 6.29 \\

     \multirow{2}{*}{$R_5$} & \multirow{2}{*}{$R_4$ + }Speed Perturb  & \multirow{2}{*}{\textbf{6.99}} & \multirow{2}{*}{\textbf{20.18}} & \multirow{2}{*}{\textbf{26.80}} & \multirow{2}{*}{\textbf{0.89}} & \multirow{2}{*}{\textbf{3.34}} & \multirow{2}{*}{\textbf{8.00}} & \multirow{2}{*}{\textbf{3.08}} & \multirow{2}{*}{\textbf{7.71}} & \multirow{2}{*}{\textbf{19.03}} & \multirow{2}{*}{\textbf{0.32}} & \multirow{2}{*}{\textbf{0.88}} & \multirow{2}{*}{\textbf{6.05}} \\
     & \text{\hspace{0.73cm}SpecAug} &   &  &  &  &  &  &  &  &  &  & &\\
    \hline

    %  $R_1$ & \text{+ SegAug} & 8.70 & 23.34 & 29.27 & 0.90 & 2.82 & 7.01 & 3.65 & 9.03 & \textbf{20.13} & 0.26 & 0.82 & \textbf{5.71} \\
    %  \hline
    %  $R_1$ & \text{Subword AED} & 8.14 & \textbf{21.88} & 29.82 & 1.07 & 2.64 & 6.98 & 3.48 & 8.95 & 20.37 & 0.47 & 0.96 & 5.90 \\
    %  $R_1$ & \text{+ SegAug} & \textbf{7.93} & 22.30 & \textbf{28.84} & \textbf{0.60} &  \textbf{1.87} & \textbf{6.23} & \textbf{3.28} & \textbf{8.64} & \textbf{19.40} & \textbf{0.27} & \textbf{0.71} & \textbf{5.47} \\
     % \hline     

    \end{tabular}
    }
\vspace{-12pt}
\end{table*}
\begin{table}[ht]
\centering
\caption{\label{tab:table2} Substitution (SUB), Deletion (DEL), and Insertion (INST) error rates (\%) on LibriSpeech and Tedlium-v3 datasets. Values in bold highlight the characteristic errors of the corresponding ASR models.}
\vspace{-4pt}
    \begin{tabular}{l  c  c  c  c  c  c}
        \hline
     \multirow{2}{*}{\textbf{\#}} & \multicolumn{3}{c}{\textbf{test other}} & \multicolumn{3}{c}{\textbf{Ted-v3}} \\
     \cline{2-7}
     & \textbf{SUB} & \textbf{DEL} & \textbf{INST} & \textbf{SUB} & \textbf{DEL} & \textbf{INST} \\
     \hline
     Subword CTC & 17.07 & 2.84 & \textbf{2.31} & 18.93 & 7.26 & \textbf{2.87} \\
     $R_3$ & 14.96 & \textbf{4.94} & 1.44 & 16.65 & \textbf{10.32} & 2.02 \\
     $R_5$ & 15.16 & 3.34 & 1.68 & 16.30 & 8.00 & 2.50 \\
    \hline
    \end{tabular}
    % \vspace{-13pt}
% \vspace{-10pt}

\end{table} 
\begin{table}[ht]
    \vspace{-5pt}
\centering
\caption{\label{tab:table3}Importance of SegAug sub-augmentation methods
on model improvement, with models trained on the small-scale dataset. Values in bold denote the worst results.}
\vspace{-2pt}
    \begin{tabular}{l  c  c  c }
        \hline
     \multirow{2}{*}{\textbf{Model}} & \multicolumn{3}{c}{\textbf{WER (\%) | DEL (\%)}} \\
     \cline{2-4}
     & \textbf{clean} & \textbf{other} & \textbf{Ted-v3} \\
     \hline
     \text{$R_5$}  & 6.99 | 0.89 & 20.18 | 3.34 & 26.80 | 8.00 \\
     \hline
      \text{- SC} & 7.54 | 1.01 & 20.75 | 3.48 & 27.59 | 8.12  \\
      \text{- SD} & \textbf{7.60} | 0.94 & \textbf{21.06} | 3.46 & 27.71 | 8.00 \\
      \text{- SM} & 7.35 | 0.97 & 20.51 | 3.47 & 27.58 | 8.23 \\
      \text{- SP} & 7.37 | \textbf{1.17} & 20.54 | \textbf{3.93} & \textbf{27.94} | \textbf{8.88} \\
    \hline
    \end{tabular}
\vspace{-12pt}
\end{table}
The results in Table \ref{tab:table1} demonstrate that standalone SegAug $(R_4)$ improves upon the non-augmented model $(R_0)$ with a relative WER reduction (rWERR) of 7.2\%, 2.7\% and 7.2\%
on test clean, test other, and Tedlium-v3, respectively. While it surpasses the results achieved with Speed Perturb, it still falls short compared to SpecAug, especially on test other.
However, the combination of SegAug with Speed Perturb and SpecAug $(R_5)$ achieves the lowest WERs across all test conditions in the small-scale dataset, outperforming other methods $(R_0$ to $R_3)$, thus highlighting the effectiveness of SegAug, especially when combined with additional augmentations. Specifically, $(R_5)$ shows a rWERR of 22.7\%, 24.8\% and 17.9\% % \(22.7_{[20.7, 24.8]}\%\), \(24.8_{[23.5, 26.0]}\%\), and \(17.9_{[16.3, 19.6]}\%\) 
over the baseline non-augmented model $(R_0)$ and further enhances the commonly used combination of SpecAug and Speed Perturb $(R_3)$, achieving an additional rWERR improvement of 12.5\%, 5.4\%, and 7.6\% on the three test sets. Notably, a substantial portion of the WER improvement is attributed to a significant reduction in the deletion (DEL) error rate. Particularly, the transition to apply SegAug from $(R_0)$ to $(R_4)$ and $(R_3)$ to $(R_5)$ yields relative DEL reductions of $(32.2\%, 22.6\%, 13.9\%)$ and $(45.4\%, 32.4\%, 22.5\%)$ across experimental test sets, respectively. These results indicate a marked decrease in deletions and underscores SegAug's impact on this type of error.

We further investigate the impact of SegAug by analyzing the distinctive characteristics of CTC and RNN-T models, with results for all error types presented in Table \ref{tab:table2}. The results clearly reflect the typical behavior of CTC and RNN-T algorithms, with CTC exhibiting higher INST and lower DEL despite the higher WER, and RNN-T showing the opposite trend. Remarkably, SegAug $(R_5)$ strikes a balance between these characteristics. 
Like CTC, SegAug demonstrates a heightened sensitivity to acoustic variations, which is evident in its significant reduction in DEL compared to $(R_3)$.
We believe that the RNN-T now emphasizes acoustic modeling with ``word-level dependency'' rather than ``sentence-level dependency'' typically introduced by the predictor. This shift enables the acoustic model to correct misspelling errors at the subword level, an issue commonly observed in CTC models \cite{zhang19g_interspeech}, while also promoting independent word assumptions within a sentence, thereby mitigating the impact of training transcription-restricted ILM.
To achieve balance, we intentionally apply SegAug with a probability of 50\% during training, allowing the RNN-T model to retain some of its sentence-level semantic coherence. This approach is reflected in the controlled INST of $(R_5)$, which is lower than that of CTC.

Additionally, we evaluate the contribution of each SegAug sub-augmentation method to model improvement by sequentially removing each one from the training configuration $(R_5)$. As shown in Table \ref{tab:table3}, the absence of any SegAug sub-augmentation results in a decline in performance, highlighting the effectiveness of each component within SegAug. Notably, eliminating SD significantly increases WER, while SP has the greatest impact on reducing DEL. This outcome aligns with the probability settings chosen for SegAug policies, as detailed in Algorithm \ref{alg:alg1}.
% \vspace{-2.5pt}
\subsection{Results on large scale dataset}
Most studies on augmentation, such as \cite{SDAFEMSR, specswap, mixspeech}, evaluate its effectiveness on small-scale datasets. Typically, augmentation methods have a significant impact on small-scale datasets, but their effect gradually lessens as the dataset size increases. For instance, as shown in Table \ref{tab:table1}, we observe only marginal improvement from Speed Perturb on large-scale data. However, both SpecAug and SegAug continue to provide meaningful benefits, demonstrating robustness in large-scale experiments.
Particularly, results on large-scale training reveal that standalone SegAug ($R_4$) reduces WER by a relative \(6.1_{[2.4, 10.1]}\%\), \(9.3_{[7.3, 11.3]}\%\), and \(6.5_{[4.7, 8.1]}\%\) compared to non-augment $R_0$ on the test clean, test other, and Ted-v3 sets, respectively. 
\begin{table}[ht]
% \vspace{-10pt}
\centering
\caption{\label{tab:table4} Evaluation of SegAug's generalization on other models.}
\vspace{-2pt}
    \begin{tabular}{l  c  c  c }
        \hline
     \multirow{2}{*}{\textbf{Model}} & \multicolumn{3}{c}{\textbf{WER (\%) | DEL (\%)}} \\
     \cline{2-4}
     & \textbf{clean} & \textbf{other} & \textbf{Ted-v3} \\
     \hline
     \text{Subword CTC}  & \textbf{3.56} | 0.28 & \textbf{8.93} | \textbf{0.80} & \textbf{20.43} | 5.92 \\
     \text{+ SegAug}  & 3.65 | \textbf{0.26} & 9.03 | 0.82 & 20.13 | \textbf{5.71} \\

     \hline
     \text{AED}  & 3.48 | 0.47 & 8.95 | 0.96 & 20.37 | 5.90 \\
     \text{+ SegAug}  & \textbf{3.28} | \textbf{0.27} & \textbf{8.64} | \textbf{0.71} & \textbf{19.40} | \textbf{5.47} \\
    \hline
    \end{tabular}
    \vspace{-13pt}
\end{table}
When combined with other augmentations ($R_5$), SegAug further reduces rWERR by \(2.5_{[-1.0, 6.9]}\%\), \(6.9_{[4.1, 9.4]}\%\), and \(4.8_{[3.1, 6.5]}\%\) over SpecAug and Speed Perturb combination ($R_3$). Additionally, ($R_5$) achieves a relative improvement over ($R_3$) in DEL of \(8.6_{[-4.7, 20.8]}\%\), \(18.5_{[9.7, 24.3]}\%\), and \(3.8_{[0.4, 6.7]}\%\) across the three test sets, demonstrating its effectiveness in large-scale training.

The confidence intervals for the results indicate that SegAug and its combinations with other augmentations generally provide consistent and statistically significant improvements in WER and DEL across various datasets, with the vast majority of outcomes showing positive gains. Remarkably, aside from the results with negative lower bounds, the remaining results consistently contain all positive values in the empirical bootstrap distribution, providing strong evidence of efficacy.
While there is a slight chance of underperformance, SegAug's enhancements remain robust and reliable in the majority of scenarios,  with the occurrence of negative values for results with lower bounds of -1.0 and -4.7 in the empirical distribution limited to only 9.28\% and 12.3\% of cases, respectively.
% \subsection{Statistical measurement and method generalization}
Additionally, SegAug's effectiveness is further highlighted by its performance on the out-of-domain Tedlium-v3 dataset. By improving the model's sensitivity to acoustic features and diversifying the learned language patterns, SegAug proves effective in enhancing speech recognition accuracy, particularly in unfamiliar conditions.

Furthermore, we assess the generalization of SegAug across other end-to-end methods, specifically CTC and AED. Table \ref{tab:table4} shows that SegAug negatively impacts CTC models. This is due to CTC's conditional independence, leading to low deletion errors, thus contradicting SegAug's goals. Conversely, SegAug demonstrates promising results on AED models, with up to 5.7\% and 42.6\% relative reductions in WER and DEL on our test sets, respectively. This effectiveness is anticipated, as AED models utilize cross-attention from textual to acoustic parts, thus implicitly learning an ILM. These results strongly support SegAug's potential as a robust augmentation technique for various speech recognition tasks.
\section{Conclusions}
In this paper, we proposed SegAug, a novel augmentation technique specifically designed to reduce deletion errors in RNN-T models for speech recognition. SegAug utilizes a combination of CTC-aligned segmentation and four sub-augmentation methods to diversify audio-text pairs and enhance the model's focus on acoustic features. Experimental results on both small and large-scale datasets demonstrated significant improvements in WER, particularly through substantial reductions in deletion errors. Moreover, SegAug showed promising generalization to other ASR models, such as AED systems, further highlighting its versatility. These findings demonstrate that SegAug is a robust and effective augmentation technique, capable of enhancing performance across a range of ASR tasks and models.

\bibliographystyle{IEEEbib}
\bibliography{refs}
\nocite{*}

\end{document}